\documentclass[journal=jacsat,manuscript=article]{achemso}

\usepackage{chemformula} % Formula subscripts using \ch{}
\usepackage[T1]{fontenc} % Use modern font encodings
\usepackage[per-mode=symbol]{siunitx}
\usepackage{xcolor}
\usepackage[nolist]{acronym}
\usepackage{subcaption}
\usepackage{graphicx}
\usepackage{graphbox} % it depends on "graphicx" package
\usepackage{placeins}
\graphicspath{{./Figures/}}

\newcommand*\polyaffi{Department of Engineering Physics, École Polytechnique de Montréal, C.P. 6079, Succ. Centre-Ville, Montréal, Québec, Canada H3C 3A7}
\newcommand{\pin}{$p$-$i$-$n$}

\newcommand{\comments}[1]{}

\usepackage[table]{xcolor}
\usepackage{tabularx}
\usepackage{booktabs}
\usepackage{siunitx}
\usepackage{amssymb} % for \checkmark
\usepackage{pifont}

\definecolor{RuleGray}{RGB}{210,210,210}

\author{A. Nomezine}
\affiliation{\polyaffi{}}
\author{G. Daligou}
\affiliation{\polyaffi{}}
\author{D. Vlassov}
\affiliation{\polyaffi{}}
\author{S. Assali}
\affiliation{\polyaffi{}}
\author{M. R. M. Atalla}
\affiliation{\polyaffi{}}
\author{O. Moutanabbir}
\email{oussama.moutanabbir@polymtl.ca}
\affiliation{\polyaffi{}}

\title{A Silicon-Compatible Uncooled Compact Broadband Infrared Spectrometer}

\keywords{American Chemical Society, \LaTeX}

\begin{document}

%%%%%%%%%%%%%%%%%%%%%%%%%%%%%%%%%%%%%%%%%%%%%%%%%%%%%%%%%%%%%%%%%%%%%
%% Definition of the different acronyms
%% 
%%%%%%%%%%%%%%%%%%%%%%%%%%%%%%%%%%%%%%%%%%%%%%%%%%%%%%%%%%%%%%%%%%%%%
%
\begin{acronym}
  \acro{HS}{heterostructure}
  \acro{DMD}{digital micromirror device}
  \acro{EVM}{Evaluation Module}
  \acro{MEMS}{microelectromechanical systems}
  \acro{GUI}{graphical user interface}
  \acro{NIR}{nearwave-infrared}
  \acro{SWIR}{shortwave-infrared}
  \acro{e-SWIR}{extended shortwave-infrared}
  \acro{MWIR}{midwave-infrared}
  \acro{SRH}{Shockley-Read-Hall}
  \acro{RP}{reduced-pressure}
  \acro{CVD}{chemical vapor deposition}
  \acro{VS}{virtual substrate}
  \acro{RIE}{reactive ion etching}
  \acro{ICP}{inductively coupled plasma}
  \acro{IPA}{isopropyl alcohol}
  \acro{BOE}{buffered oxide etch}
  \acro{PECVD}{plasma-enhanced chemical vapor deposition}
  \acro{RPCVD}{reduced pressure chemical vapor deposition}
  \acro{VF}{view factor}
  \acro{SLS}{stabilized light source}
  \acro{FTIR}{Fourier transform infrared}
  \acro{IR}{infrared}
  \acro{XRD}{X-ray diffraction}
  \acro{RSM}{reciprocal space mapping}
  \acro{TEM}{transmission electron micrograph}
  \acro{APT}{atom probe tomography}
  \acro{GeSn}{Germanium Tin}
  \acro{EDX}{Energy-dispersive X-ray}
  \acro{RT}{room temperature}
\end{acronym}

% \begin{tocentry}

% Some journals require a graphical entry for the Table of Contents.
% This should be laid out ``print ready'' so that the sizing of the

% Inside the \texttt{tocentry} environment, the font used is Helvetica
% 8\,pt, as required by \emph{Journal of the American Chemical

% The surrounding frame is 9\,cm by 3.5\,cm, which is the maximum
% permitted for  \emph{Journal of the American Chemical Society}
% graphical table of content entries. The box will not resize if the
% content is too big: instead it will overflow the edge of the box.

% This box and the associated title will always be printed on a
% separate page at the end of the document.

% \end{tocentry}

\begin{abstract}

Infrared spectroscopy is a widely used technique for molecular identification, yet its widespread deployment remains constrained by bulky instrumentation, cryogenic cooling requirements, and limited portability. Here we demonstrate broadband near- to short-wave infrared spectroscopy enabled by a silicon-compatible GeSn photodetector integrated into a compact digital micromirror device-based single-pixel architecture. By combining detector design with targeted optical reconfiguration and wavelength recalibration, we extend the operational range up to \SI{2.4} {\micro\meter} while maintaining room-temperature, zero-bias photovoltaic operation. System-level benchmarking quantifies the intrinsic trade-off between spectral extension and detectivity in narrow-bandgap photodetection, yielding a specific detectivity of $1.12\times10^{10}$~cm\,Hz$^{1/2}$\,W$^{-1}$ at \SI{1.55}{\micro\meter} with an extended cutoff wavelength. We validate the capabilities of the obtained spectrometer through selective identification of representative commodity plastics, demonstrating a lightweight, low-power, and mechanically robust platform for portable and autonomous deployment. Despite the expected increase in generation-limited noise, access to strong polymer combination bands beyond \SI{1.7}{\micro\meter} allows enhanced material discrimination. This work establishes a pathway toward scalable distributed molecular sensing and broadens access to infrared spectroscopy for environmental monitoring, industrial process control, and community-level chemical detection.

\end{abstract}

\section{INTRODUCTION}

Infrared (IR) spectroscopy provides a universal, label-free means of identifying molecules through their vibrational fingerprints and underpins technologies across environmental monitoring, healthcare, industry, and security\cite{Siesler2008-cl,Gupta2022-mo,bakerUsingFourierTransform2014,bellisolaInfraredSpectroscopyMicroscopy2011,shepherdInfraredSpectroscopyEnabling2007,Caygill2012-ok}. However, despite its conceptual simplicity and analytical power, access to IR spectroscopy remains largely restricted to specialized facilities and expert users. Indeed, state-of-the-art instruments are typically confined to laboratory settings, relying on bulky optics, cryogenically cooled detectors, and mechanically complex architectures\cite{Guan2023-pg,Zou2023-vu}. This paradigm has created a structural gap between the ability to measure and the urgency to act, particularly in contexts where rapid, distributed, and localized chemical sensing is essential.

Democratizing IR spectroscopy, by making it compact, affordable, and scalable, would enable communities, industries, and autonomous systems to directly interrogate their chemical environments in real time \cite{Yan2023-zi, Cebi2023-ae,Daham2005-uf,Li2023-ah,Zhao2023-wl}. This transition would have profound implications by enabling communities to detect environmental pollutants and toxins without relying on distant analytical facilities, achieve rapid response to contamination events, and support continuous monitoring of ecosystems under stress\cite{Ateia2024-sc,Heo2023-jk}. The demand for such capabilities is particularly acute across a range of emerging applications. For instance, detection of persistent pollutants remains a major global challenge, requiring sensitive, field-deployable analytical tools\cite{Trinh2024-uv}. Near-infrared (NIR) spectroscopy is widely employed in chemical and biological sensing, where overtone and combination vibrations of C--H, O--H, and C--O bonds provide material-specific spectral fingerprints \cite{Taneepanichskul2022-lt,Workman2007-vn,Scheirs1998-cx}. While conveyor-based systems enable high-throughput automated sorting, compact and field-deployable spectrometers are increasingly relevant for decentralized screening. While many commercially available compact NIR spectrometers remain limited to wavelengths below \SI{1.7}{\micro\meter}, recent computational single-pixel approaches have begun to extend spectral coverage beyond this conventional limit. For instance, Liang et al. demonstrated a miniaturized spectrometer combining a voltage-tunable liquid-crystal/Au modulation stack with a PbS photodetector, achieving a tunable spectral response spanning \SIrange{1.15}{2.0}{\micro\meter} and enabling computational spectral reconstruction and plastic classification through sequential bias-dependent measurements \cite{Liang2025-qc}. However, the plastic spectral reconstruction experiments remain within the \SIrange{1.25}{1.75}{\micro\meter} range, while the sequential voltage scanning impose limitations on acquisition speed. Extending compact spectroscopy further into the eSWIR is particularly desirable since this region contains stronger and more distinctive molecular absorption features that can enhance material separability \cite{Scheirs1998-cx}. Achieving such spectral coverage in a low-power, field-deployable system while maintaining room-temperature, zero-bias operation and compatibility with silicon manufacturing remains a major system-level challenge.

Compact spectrometers with extended IR operation range integrated into unmanned systems or distributed sensor networks could provide continuous, spatially resolved chemical mapping in scenarios ranging from wildfire monitoring to hazardous material detection\cite{Allison2016-ag,Ong2019-hv}. Meeting these needs imposes stringent and often competing requirements: operation at room temperature, low power consumption, minimal size and weight, mechanical robustness, and cost-effectiveness, all while maintaining meaningful spectral resolution and sensitivity. Crucially, these systems must also be manufacturable on a scale to enable widespread adoption beyond niche or high-cost applications.

By exploiting emerging group IV GeSn semiconductors \cite{Moutanabbir2021}, here we address these limitations and demonstrate a compact, cost-effective, and silicon-compatible infrared spectrometer that advances this vision of distributed molecular sensing. The developed system consists of a lightweight and robust architecture potentially suitable for deployment in portable and autonomous platforms. We also introduce this compact spectrometer to demonstrate selective molecular identification across representative use cases of commodity plastic, highlighting the potential of this approach for real-time environmental monitoring, industrial process control, and community-level detection of chemicals.

\section{EXPERIMENTAL DETAILS}

The photodetectors investigated in this work were grown epitaxially on silicon wafers following the protocol described in the Methods section. Figure~\ref{fig1}a illustrates the Ge$_{0.91}$Sn$_{0.09}$ \pin{} photodiode heterostructure, comprising a Ge virtual substrate and a compositionally graded GeSn buffer layers that mitigate the lattice mismatch between Si and the active GeSn device layers. The device adopts a vertical double-mesa geometry, with extended short-wave infrared (eSWIR) absorption enabled by the narrow bandgap of the $9\%$ Sn GeSn i-layer. Structural and growth characteristics are detailed in the Methods section. An optical micrograph of the fabricated 1~mm-diameter device is shown in Fig.~\ref{fig1}b.

For system-level evaluation, the Ge$_{0.91}$Sn$_{0.09}$ \pin{} photodiode was integrated into a commercial TO-18 metal-can package, providing a compact and mechanically robust interface compatible with the portable spectrometer platform. The device was mounted using conductive silver paste to establish backside electrical contact, while the top p- and n-contacts were wire-bonded to the package leads, preserving vertical carrier transport through the \pin{} junction and minimizing series resistance. A photograph of the packaged 1~mm device is shown in Fig.~\ref{fig1}c.

The packaging strategy preserves optical access through a reinstalled anti-reflection window, enabling efficient photon coupling in the eSWIR range. Care was taken to minimize parasitic capacitance and leakage pathways introduced by the housing, allowing stable room-temperature operation under zero-bias photovoltaic conditions. No measurable degradation in dark current or responsivity was observed after packaging, indicating that the integration process does not introduce significant additional surface or contact-related recombination.

\section{RESULTS AND DISCUSSION}
\FloatBarrier
\subsection{Detector performance and system-level trade-offs}

\begin{figure}[!htb] 
\centering \includegraphics[width=\textwidth]{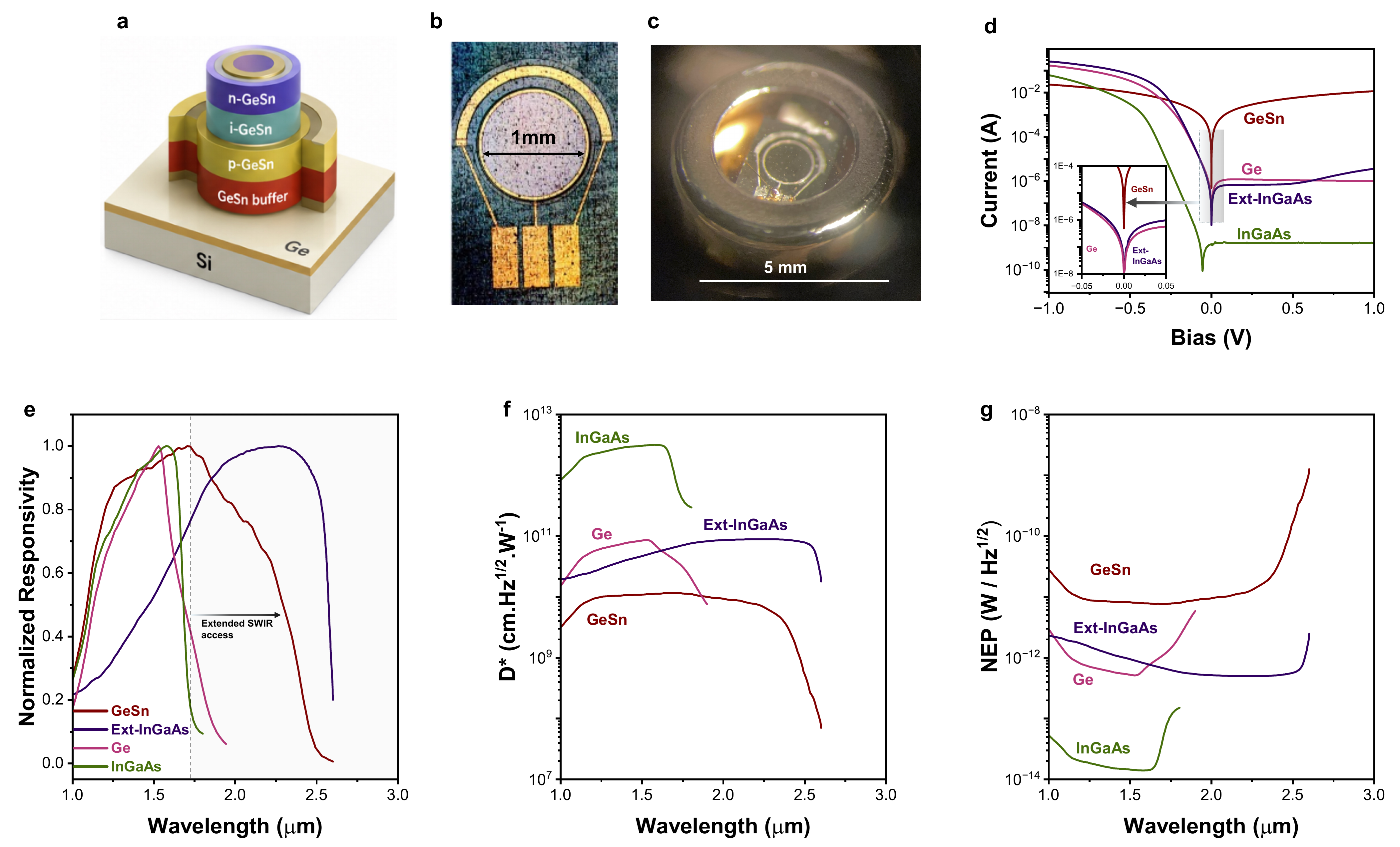} 
\caption{%
Structural design, device realization, and performance benchmarking of a 1-mm-diameter GeSn \textit{p--i--n} photodiode. 
(a) Three-dimensional schematic of the GeSn double-mesa architecture grown on Si, illustrating the Ge virtual substrate, graded GeSn buffer, and vertical \textit{p--i--n} stack. 
(b) Optical micrograph of the fabricated mesa device prior to packaging. 
(c) Photograph of the wire-bonded photodiode mounted in a TO18-can package. 
(d) Dark current–voltage characteristics at 293 K measured from $-1$ to $+1$ V, compared with commercial InGaAs (1 mm), extended InGaAs (1 mm), and Ge (0.5 mm) photodiodes. 
(e) Normalized spectral responsivity, highlighting the long-wavelength cut-off extension enabled by GeSn relative to Ge. 
(f) Specific detectivity ($D^*$) at 293 K as a function of wavelength for the same detector set. 
(g) Noise-equivalent power (NEP) at 293 K extracted from measured responsivity and dark current.
}
\label{fig1} 
\end{figure}

To benchmark performance in the spectrometer context, four sets of photodetectors were characterized under identical electrical operation (0~V photovoltaic mode) and consistent optical coupling: a standard In$_{0.53}$Ga$_{0.47}$As photodiode (1~mm; Hamamatsu G12180-010A), a Ge photodiode (0.5~mm; GPD Optoelectronics GM03), an extended In$_{0.80}$Ga$_{0.20}$As photodiode (1~mm; Thorlabs FD10D), and the Ge$_{0.91}$Sn$_{0.09}$ \pin{} photodiode developed here (1~mm). Zero-bias operation corresponds to the native configuration of the spectroscopic system platform and avoids bias-induced gain mechanisms and associated excess noise, enabling a direct comparison of intrinsic material-limited performance.

Table~\ref{Table1} summarize the electrical characteristics at 293~K. As reported in Figure~\ref{fig1}d, Ge$_{0.91}$Sn$_{0.09}$ exhibits a higher dark current ($I_{\mathrm{dark}}=4.90\times10^{-7}$~A) than Ge ($1.33\times10^{-8}$~A) at 0 V bias voltage. This behavior is consistent with the reduced bandgap ($E_g \approx 0.51$~eV at room temperature, as extracted from photoluminescence measurements) and with defect-assisted generation pathways associated with partial strain relaxation in GeSn heteroepitaxial layers, which enhance Shockley--Read--Hall recombination at room temperature\cite{Giunto2024-tq,Atalla2023-xp}. The corresponding shunt resistance decreases from $2.03\times10^{4}$~$\Omega$ (Ge) to $2.94\times10^{3}$~$\Omega$ (GeSn), increasing the generation-limited noise. These trends translate into a decrease in the device detectivity. At \SI{1.55}{\micro\meter}, the specific detectivity decreases from $8.32\times10^{10}$~cm\,Hz$^{1/2}$\,W$^{-1}$ (Ge) to $1.12\times10^{10}$~cm\,Hz$^{1/2}$\,W$^{-1}$ (GeSn), remaining within one order of magnitude under zero-bias operation. Consistently, the NEP increases from $5.32\times10^{-13}$ to $7.95\times10^{-12}$~W\,Hz$^{-1/2}$ (Table~\ref{Table1}). Despite the associated increase in generation-limited noise, the GeSn photodiode provides a distinct advantage in spectral reach. As shown in Fig.~\ref{fig1}e, the normalized responsivity extends to approximately \SI{2.4}{\micro\meter}, whereas Ge devices are intrinsically limited to \SI{\sim1.8}{\micro\meter}.

A comparable trade-off is observed for the III--V platform when transitioning from standard to extended InGaAs. Extending the spectral cut-off from \SI{1.7}{\micro\meter} to \SI{2.6}{\micro\meter} (Fig.~\ref{fig1}e) is accompanied by a rise in the zero-bias dark current from $6.35\times10^{-11}$~A to $1.04\times10^{-8}$~A and a decrease in shunt resistance from $8.9\times10^{7}$~$\Omega$ to $3.79\times10^{4}$~$\Omega$ (Table~\ref{Table1}). At \SI{1.55}{\micro\meter}, $D^*$ decreases from $3.14\times10^{12}$ to $5.8\times10^{10}$~cm\,Hz$^{1/2}$\,W$^{-1}$ (nearly two orders), while NEP increases from $1.41\times10^{-14}$ to $8.55\times10^{-13}$~W\,Hz$^{-1/2}$ (Fig.~\ref{fig1}f,g). Although bandgap reduction introduces an additional intrinsic noise in both material systems, these results indicate that the degradation associated with the spectral extension is relatively less severe for the group IV systems thus offering a balanced compromise between spectral coverage and intrinsic noise within a silicon-compatible architecture.

\newcolumntype{Y}{>{\centering\arraybackslash}X}

% Subtle styling (Nature-friendly)
\definecolor{RuleGray}{RGB}{210,210,210}
\definecolor{GroupGray}{RGB}{252,252,252}

% siunitx setup for aligned scientific notation in tables
\sisetup{
  detect-weight = true,
  detect-family = true,
  exponent-product = \times,
  table-number-alignment = center,
  table-text-alignment = center,
  inter-unit-product = \cdot
}

% ---------- Ultra-polished table ----------
\begin{table}[t]
\centering
\caption{Electrical and optical performance benchmarking of the investigated photodetectors at 293~K under zero-bias photovoltaic operation.}
\label{Table1}

\renewcommand{\arraystretch}{1.18}
\setlength{\tabcolsep}{6pt}

\begin{tabularx}{\textwidth}{lYYYY}
\arrayrulecolor{RuleGray}\toprule
 & InGaAs & Ge & Ext-InGaAs & \textbf{Ge$_{0.91}$Sn$_{0.09}$} \\
\arrayrulecolor{RuleGray}\midrule

\rowcolor{GroupGray}
\multicolumn{5}{l}{\textbf{\textit{Electrical characteristics (0~V)}}}\\
$R_{\mathrm{shunt}}$ (\si{\ohm})
& \num{8.90e7} & \num{2.03e4} & \num{3.79e4} & \num{2.94e3} \\
$I_{\mathrm{dark}}$ (\si{\ampere})
& \num{6.35e-11} & \num{1.33e-8} & \num{1.04e-8} & \num{4.9e-7} \\
$I_{\mathrm{therm}}$ (\si{\ampere})
& \num{1.35e-14} & \num{8.9e-13} & \num{6.0e-13} & \num{2.34e-12} \\
$I_{\mathrm{shot}}$ (\si{\ampere})
& \num{2.15e-14} & \num{6.5e-14} & \num{5.9e-14} & \num{3.96e-13} \\

\addlinespace[4pt]
\arrayrulecolor{RuleGray}\midrule

\rowcolor{GroupGray}
\multicolumn{5}{l}{\textbf{\textit{Optical performance at 1.55 \si{\micro\meter}}}}\\
$D^{*}$ (\si{cm.Hz^{1/2}.W^{-1}})
& \num{3.14e12} & \num{8.32e10} & \num{5.8e10} & \num{1.12e10} \\
NEP (\si{W.Hz^{-1/2}})
& \num{1.41e-14} & \num{5.32e-13} & \num{8.55e-13} & \num{7.95e-12} \\
$R$ (\si{A.W^{-1}})
& 1.79 & 1.67 & 0.77 & 0.32 \\

\addlinespace[4pt]
\arrayrulecolor{RuleGray}\midrule

\rowcolor{GroupGray}
\multicolumn{5}{l}{\textbf{\textit{Spectral limit}}}\\
Cut-off $\lambda$ (\si{\micro\meter})
& 1.7 & 1.8 & 2.6 & 2.4 \\

\arrayrulecolor{RuleGray}\bottomrule
\end{tabularx}

\arrayrulecolor{black} % reset for later tables
\end{table}

\FloatBarrier
\subsection{Spectrometer hardware development}

\begin{figure}[!htb]
  \centering
  \includegraphics[width=1\textwidth]{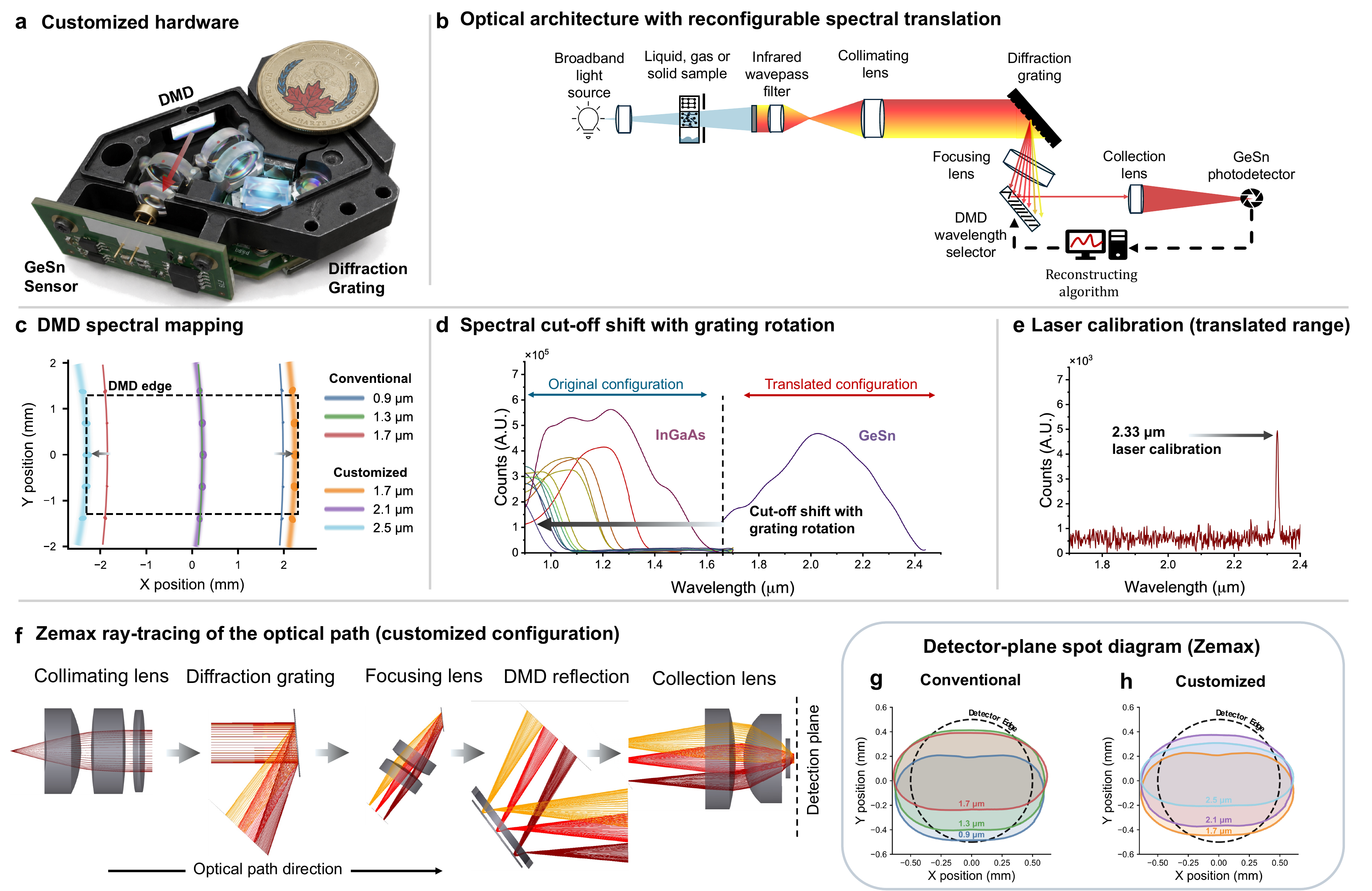}
  \caption{Optical reconfiguration of the compact spectrometer platform for SWIR operation using GeSn detection. 
(a) Three-dimensional rendering of the customized NIRscan Nano spectrometer integrating the GeSn photodetector, diffraction grating, and digital micromirror device (DMD), shown relative to a Canadian \textbf{\$}1 coin for scale. 
(b) Schematic of the modified optical architecture enabling spectral translation, while preserving DMD-based wavelength selection and single-pixel detection. 
(c) Simulated wavelength mapping onto the DMD plane for the original and customized configurations, illustrating translation of the accessible spectral window toward the SWIR range. 
(d) Measured InGaAs detector response as a function of wavelength for different grating positions, demonstrating progressive spectral translation from the original \SIrange{0.9}{1.7}{\micro\meter} configuration toward the customized \SIrange{1.7}{2.5}{\micro\meter} window. 
(e) Experimental wavelength calibration using a monochromatic \SI{2.33}{\micro\meter} laser source, confirming an effective spectral translation of \SI{740}{\nano\meter} relative to the original spectral mapping without modification of instrument firmware. 
(f) Zemax ray-tracing simulations of the customized optical configuration, showing preservation of beam collimation, diffraction, focusing, DMD reflection, and collection through the translated spectral range. 
(g,h) Zemax detector-plane spot diagrams for the (g) original and (h) customized configurations, confirming that wavelength-dependent beam profiles remain confined within the detector active area across the explored spectral windows. 
}
  \label{fig2}
\end{figure}

Figure~\ref{fig2}a illustrates the customized spectrometer hardware integrating the GeSn photodetector, while Fig.~\ref{fig2}b shows the modified optical architecture combining diffraction-grating dispersion, digital micromirror device (DMD) wavelength selection, and single-pixel detection. The compact spectrometer platform was reconfigured to extend operation to eSWIR range while preserving the single-pixel architecture (Fig.~\ref{fig2}a--b). 

To access the eSWIR spectral window, the conventional near-infrared band-pass filter was replaced with a long-pass filter featuring a cut-on wavelength of \SI{1.65}{\micro\meter}, enabling transmission beyond the intrinsic InGaAs cut-off. A plane-ruled grating optimized with blaze wavelength of \SI{2}{\micro\meter} was introduced instead of the diffraction grating, allowing efficient first-order diffraction across the targeted \SIrange{1.7}{2.5}{\micro\meter} range while preserving compact optical geometry.

Simulated wavelength mapping onto the DMD plane confirms translation of the accessible spectral window toward longer wavelengths while maintaining compatibility with the fixed DMD geometry (Fig.~\ref{fig2}c). The translated eSWIR configuration produces a slightly broader and laterally displaced spectral footprint on the DMD compared to that of the conventional NIR configuration. This effect is attributed primarily to the chromatic response of the focusing lens assembly, which was originally optimized for the \SIrange{0.9}{1.7}{\micro\meter} range, rather than to a loss of grating dispersion control. The calculated displacement remains limited to approximately \SI{0.5}{\milli\meter} at the spectral edges, while the central wavelengths remain closely aligned with the original DMD mapping. Importantly, the full translated spectral range remains incident on the active DMD area, allowing wavelength selection and reconstruction to be preserved.

The spectral translation was experimentally verified by monitoring the detector response as a function of the grating position (Fig.~\ref{fig2}d). Using the original InGaAs detector and direct broadband illumination, rotation of the diffraction grating progressively displaced the apparent InGaAs cut-off across the reconstructed wavelength axis, providing an experimental marker of the translated wavelength scale. In the original configuration, measurable signal is restricted near the intrinsic InGaAs cut-off around \SIrange{1.65}{1.70}{\micro\meter}. Following optical reconfiguration for the GeSn photodetector, the measurable response extends into the eSWIR, reaching approximately \SI{2.5}{\micro\meter}. A monochromatic \SI{2.33}{\micro\meter} laser source was then used as an absolute spectral marker, confirming the extended detection range (Fig.~\ref{fig2}e).

Zemax ray-tracing simulations further verified the preservation of the optical path in the customized configuration (Fig.~\ref{fig2}f). The extended operating range induces a small longitudinal focal shift relative to the original DMD plane, consistent with the chromatic response of the lens assembly. Nevertheless, the simulated beam remains sufficiently collimated and correctly directed through the diffraction, focusing, DMD reflection, and collection stages. Detector-plane spot diagrams calculated under the same field conditions as the reference configuration confirm that wavelength-dependent beam profiles remain confined within the detector active area in both configurations (Fig.~\ref{fig2}g--h). This confirms that the reconfigured optical path remains compatible with single-pixel collection in the eSWIR range.

In the customized configuration, the effective spectral resolution is governed by DMD sampling of the dispersed spectrum rather than by the intrinsic grating resolving power. Based on DMD sampling and reconstruction constraints, the effective resolution is approximately \SI{10}{\nano\meter} across the \SIrange{1.7}{2.5}{\micro\meter} range (Supplementary Information). Operation remained restricted to the first diffraction order, with no observable spectral overlap within the explored spectral window.

\FloatBarrier
\subsection{Applications: NIR versus eSWIR spectroscopy}

\begin{figure}[!htp]
  \centering
  \includegraphics[width=1\textwidth]{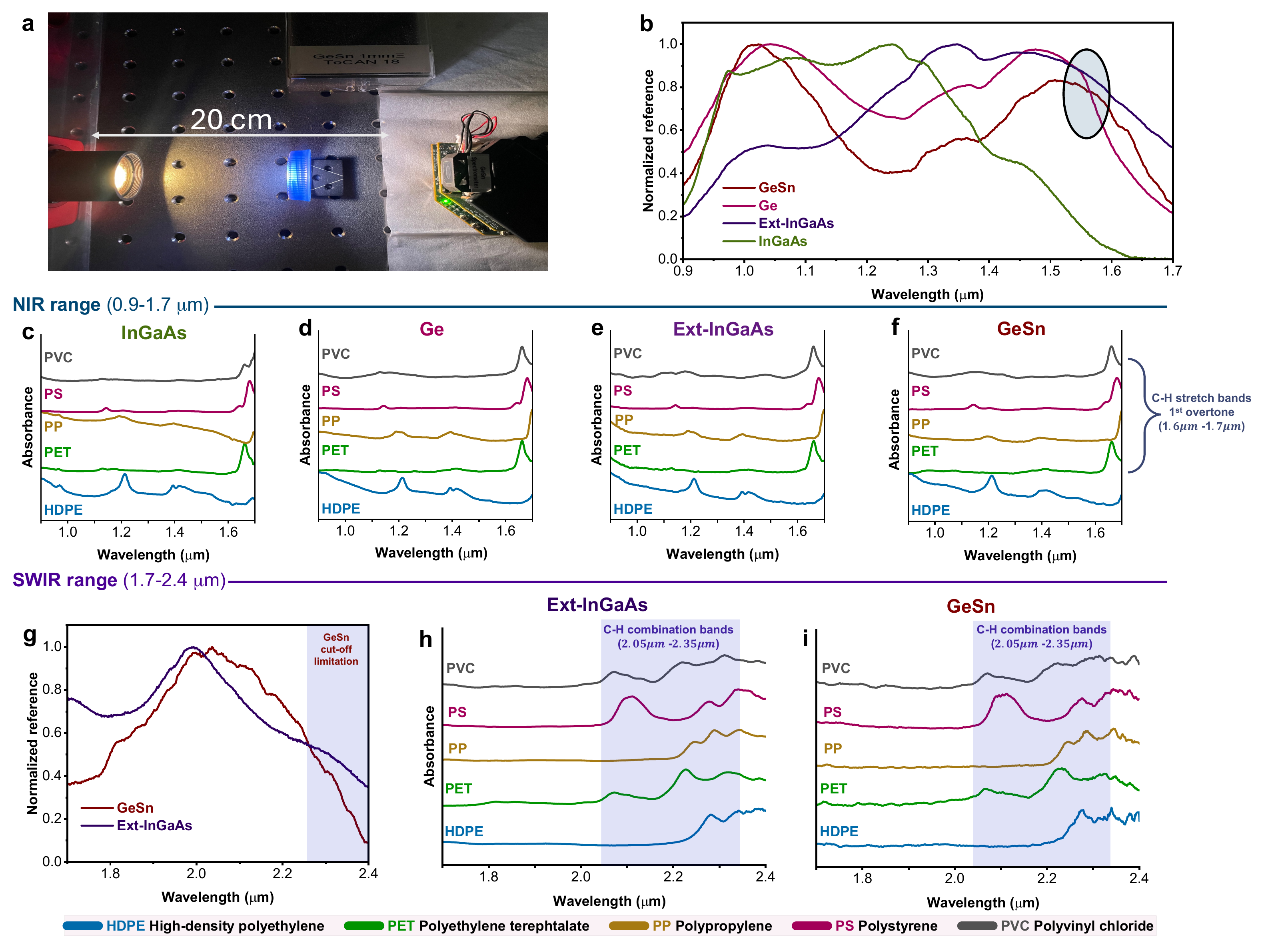}
\caption{Customized spectroscopy to the \SIrange{1.7}{2.4}{\micro\meter} range enabled by optical reconfiguration and GeSn detection.
(a) Photograph of the polymers spectrum acquisition.
(b) Normalized spectral response of the InGaAs, Ge, GeSn \textit{p--i--n}, and extended InGaAs photodetectors used in this study under identical illumination conditions. 
(c–f) Absorption spectra of polyvinyl (PVC), polystyrene (PS), polypropylene (PP), polyethylene terephthalate (PET), and high-density polyethylene (HDPE) measured over the \SIrange{0.9}{1.7}{\micro\meter} wavelength range using (c) InGaAs, (d) Ge, (e) extended InGaAs, and (f)  GeSn \textit{p--i--n} photodetectors. 
(g) Normalized reference signal of the ext-InGaAs and GeSn \textit{p--i--n} photodetectors in the extended spectral configuration, demonstrating broadband
sensitivity up to 2.4 µm.
(h) Absorption spectra acquired with an extended InGaAs photodetector over the 1.7 µm to 2.4 µm range. 
(i) Corresponding absorption spectra of PVC, PS,
PP, PET, and HDPE measured using the GeSn \textit{p--i--n} photodetector}
\label{fig3}
\end{figure}

As a proof-of-principle operation of the obtained broadband spectrometer, we conducted systematic spectroscopic studies focused on real-time plastic sorting. All measurements were performed using a stabilized broadband tungsten--halogen illumination source coupled through a collimating lens positioned approximately \SI{20}{\centi\meter} from the spectrometer entrance aperture. The same optical train, DMD pattern sequence, and Hadamard-based reconstruction algorithm were used for the InGaAs, Ge, extended InGaAs, and GeSn photodetectors to ensure direct system-level comparability. 

Figure~\ref{fig3}a shows the experimental configuration used for polymer spectroscopic measurements. Polymer samples (HDPE, PP, PET, PVC, and PS) were measured in transmission at room temperature with all photodetectors operated under zero-bias photovoltaic conditions. Absorption spectra were normalized to compare spectral features and band contrast independently of absolute detector gain.

Figure~\ref{fig3}b presents the normalized reference signals acquired under identical illumination conditions for the InGaAs, Ge, GeSn, and extended InGaAs photodetectors integrated into the original spectrometer platform. These curves represent the overall system response, including source spectral distribution, optical throughput, diffraction efficiency, and detector sensitivity, rather than the intrinsic detector responsivity alone. Within the \SIrange{0.9}{1.7}{\micro\meter} spectral window, all detectors exhibit comparable signal coverage, confirming the consistent operation of the spectrometer platform across detector technologies. The normalized responses of Ge and GeSn photodetectors closely overlap throughout most of this range, reflecting their similar spectral behavior below the intrinsic Ge absorption edge. Beyond approximately \SI{1.6}{\micro\meter}, however, the GeSn response exceeds that of Ge as the latter approaches its intrinsic cut-off.

Absorption spectra acquired in the original NIR configuration are presented in Fig.~\ref{fig3}c--f. Dominant absorption features appear between \SIrange{1.2}{1.45}{\micro\meter}, corresponding to second C--H stretching vibrations and C--H combination overtones characteristic of commodity polymers \cite{Wu2020-le}. Additional spectral structure emerges near the long-wavelength edge of the accessible range (\SIrange{1.6}{1.7}{\micro\meter}), particularly for PS, PET, PVC, and PP. In the standard InGaAs configuration, several of these bands lie near the intrinsic detector cut-off and are partially truncated, limiting full access to the long-wavelength overtone region.

Comparable peak positions are observed using Ge, GeSn, and extended InGaAs photodetectors, confirming preservation of wavelength mapping following detector substitution. The HDPE feature near \SI{1.4}{\micro\meter} retains its characteristic double structure in the InGaAs, Ge, and extended InGaAs measurements, whereas the GeSn spectrum appears slightly smoother owing to increased spectral binning implemented to compensate for reduced signal-to-noise ratio. Importantly, diagnostically relevant bands near \SI{1.7}{\micro\meter} remain clearly resolved using the GeSn photodetector, including the principal PVC and PP features that are partially truncated in standard InGaAs measurements.

Using the modified optical configuration, polymer spectra were subsequently acquired over the \SIrange{1.7}{2.4}{\micro\meter} range using both GeSn and extended InGaAs photodetectors. Figure~\ref{fig3}g presents the normalized reference signals obtained in the extended spectral configuration. The GeSn photodetector maintains a comparatively uniform response throughout the \SIrange{2.0}{2.2}{\micro\meter} region, followed by a gradual decrease as the detector approaches its intrinsic cut-off near \SI{2.4}{\micro\meter}. Figures~\ref{fig3}h--i show the corresponding absorption spectra of the investigated polymers measured in the eSWIR range. In this spectral region, strong C--H combination bands emerge between \SIrange{2.05}{2.35}{\micro\meter}, providing substantially higher spectral contrast than the overtone features observed below \SI{1.7}{\micro\meter}. Under identical normalization conditions, the peak-to-baseline contrast of the dominant polymers band near \SI{2.3}{\micro\meter} increases by approximately a factor of 2--3 relative to the strongest overtone feature near \SI{1.6}{\micro\meter}. HDPE and PP exhibit pronounced multi-peak structure between \SIrange{2.25}{2.35}{\micro\meter}, with polymer-specific differences in relative peak intensity and peak separation, while PET and PVC display distinctive spectral signatures throughout the \SIrange{2.1}{2.3}{\micro\meter} region. The spectra obtained using the GeSn photodetector closely reproduce those acquired with extended InGaAs, confirming accurate wavelength mapping and comparable spectral fidelity across the eSWIR range.

Table~\ref{Tab2} summarizes the accessibility of characteristic polymer absorption bands for each detector platform. Whereas standard InGaAs and Ge photodetectors are fundamentally restricted to overtone features below approximately \SI{1.7}{\micro\meter}, both GeSn and extended InGaAs provide access to chemically specific SWIR combination bands that remain inaccessible in conventional compact spectrometers.

Despite the reduced detectivity, the GeSn photodetector maintains sufficient signal-to-noise ratio to resolve polymer absorption features throughout the \SIrange{1.7}{2.4}{\micro\meter} spectral window. These results demonstrate that, in compact single-pixel spectrometers, practical material discrimination is governed primarily by access to high-contrast combination bands in the eSWIR range rather than by maximizing detectivity within the conventional NIR range alone. By extending the operational bandwidth to \SI{2.4}{\micro\meter} within a silicon-compatible and zero-bias architecture, the GeSn-enabled platform unlocks chemically informative spectral signatures that substantially enhance polymer identification in a compact, field-deployable format.

\begin{table*}[!htb]
\centering
\small
\setlength{\tabcolsep}{4pt}
\renewcommand{\arraystretch}{1.05}
\caption{Summary of characteristic polymer absorption bands\cite{Weyer2006-ih} within the investigated spectral range and detector accessibility under the original and extended configurations. 
Checkmarks indicate clearly resolved peaks, tildes indicate partial truncation or reduced contrast, and crosses indicate absence due to spectral cut-off. 
\checkmark: clearly resolved; $\sim$: partially resolved or reduced contrast; $\times$: not accessible within spectral window.}
\label{Tab2}

\renewcommand{\arraystretch}{1.15}
\setlength{\tabcolsep}{6pt}

\begin{tabular}{llllcccc}
\toprule
Polymer & Peak (nm) & Vibrational assignment 
& InGaAs & Ge & \textbf{GeSn} & Ext-InGaAs \\
\midrule

PP   & 1180-1250 & 3$\nu$(C-H) (second overtone) \cite{Mozaffari2026-sv,Cucuzza2026-vk}      
     & \checkmark & \checkmark & \checkmark & \checkmark \\

PP   & 1380-1420 & Combination bands (CH$_3$ / CH$_2$)\cite{Mozaffari2026-sv,Cucuzza2026-vk}       
     & \checkmark & \checkmark & \checkmark & \checkmark \\

PP   & 1680-1750 & 2$\nu$(asymmetric C-H stretch) (1st overtone) \cite{Mozaffari2026-sv,Cucuzza2026-vk}
     & $\times$ & $\sim$ & \checkmark & \checkmark \\

PP   & 2280--2330 & C-H stretch + bend (comb.) 
     & $\times$ & $\times$ & \checkmark & \checkmark \\

HDPE & 1210 & 3$\nu$(C-H stretch) (second overtone) \cite{Cucuzza2026-vk}
     & \checkmark & \checkmark & \checkmark & \checkmark \\

HDPE & 1390-1420 & Combination bands (CH$_3$ / CH$_2$)    
     & \checkmark & \checkmark & $\sim$ & \checkmark \\

HDPE & 2300--2350 & CH stretch + bend (comb.)    
     & $\times$ & $\times$ & \checkmark & \checkmark \\

PET  & 1660 & 2$\nu$(C-H stretch) (1st overtone) \cite{raniMiniaturizedNearinfraredMicroNIR2019}         
     & \checkmark & \checkmark & \checkmark & \checkmark \\

PET  & 2100--2200 & (C–H / C=O) Combination region
     & $\times$ & $\times$ & \checkmark & \checkmark \\

PS   & 1142 & 3$\nu$(aromatic C-H stretch) (second overtone) \cite{Cucuzza2026-vk}                   
     & \checkmark & \checkmark & \checkmark & \checkmark \\

PS   & 1700 & 2$\nu$(aromatic C-H stretch) (1st overtone) \cite{Cucuzza2026-vk}
     & $\sim$ & \checkmark & \checkmark & \checkmark \\

PS   & 2050--2150 & Combination bands (aromatic C–H) \cite{Cucuzza2026-vk}
     & $\times$ & $\times$ & \checkmark & \checkmark \\

PVC  & 1700 & 2$\nu$(C-H) (1st overtone)        & $\sim$ & \checkmark & \checkmark & \checkmark \\

PVC  & 2100--2300 & Combination bands   
     & $\times$ & $\times$ & \checkmark & \checkmark \\

\bottomrule
\end{tabular}
\end{table*}

\FloatBarrier
\section{CONCLUSION}
This work demonstrates a system-level route to extend compact spectroscopy by integrating a silicon-compatible GeSn \pin{} photodiode within a MEMS-based spectrometer architecture. Quantitative benchmarking confirms the expected trade-off associated with bandgap reduction: extending the spectral response to \SI{2.4}{\micro\meter} increases generation-limited dark current and reduces detectivity in the conventional \SI{1.55}{\micro\meter} band. However, the degradation remains within one order of magnitude relative to Ge under zero-bias operation, while enabling access to wavelength regions that are inaccessible to standard InGaAs.

The practical significance of this trade-off emerges at the system level. Polymer discrimination in compact spectrometers is governed not only by peak detectivity but by access to chemically specific absorption bands. The \SIrange{1.7}{2.4}{\micro\meter} window contains strong C--H combination bands that exhibit higher contrast and richer multi-peak structure than overtone features below \SI{1.7}{\micro\meter}. By extending spectral coverage into this region, the GeSn-enabled platform improves material separability despite its lower near-infrared detectivity, demonstrating that spectral reach can outweigh detectivity penalties in application-driven architectures.

Compared with extended InGaAs, which achieves long-wavelength response through increased indium content and metamorphic strain management, GeSn offers a group-IV pathway that is intrinsically aligned with silicon-based integration. While both approaches incur elevated dark current upon bandgap reduction, the detectivity degradation relative to spectral extension is comparatively moderate in the GeSn platform under photovoltaic operation. This distinction, together with CMOS compatibility and monolithic integration prospects, positions GeSn as a scalable alternative for broadband SWIR detection in compact systems.

Overall, these results show that broadband extension to \SI{2.4}{\micro\meter} within a zero-bias, silicon-compatible architecture is feasible without prohibitive loss of functional performance. By coupling group-IV bandgap engineering with compact MEMS-based spectral selection, the demonstrated approach establishes a pathway toward scalable, field-deployable SWIR spectrometers optimized for application-specific chemical sensing.

\FloatBarrier
\section{Methods}

\FloatBarrier
\subsection{Growth of GeSn epilayers}

The GeSn \pin{} heterostructures investigated here were grown on 4-inch Si(100) substrates by reduced-pressure chemical vapor deposition, following the layer sequence and growth conditions reported in Ref.~\cite{Atalla2024-yj}. The epitaxial stack comprises a Ge virtual substrate and a compositionally graded GeSn buffer engineered to accommodate the lattice mismatch associated with Sn incorporation by confining dislocations to the lower buffer region\cite{Hartmann2025-np,Assali2018-xw,Assali2019-wg,Taoka2016-ew,Giunto2024-ay,Gallagher2015-wu,Aubin2017-bx,Senaratne2016-zu,Dou2018-cc,Dou2018-zi,Castioni2022-ew}.

The active region is a vertical GeSn \pin{} junction incorporating an intrinsic layer with $\sim$9~at.\% Sn, yielding a reduced effective bandgap and infrared absorption extending beyond the intrinsic Ge cut-off (Fig.~\ref{fig1}e). High doping in the p- and n-type contact layers supports low series resistance and efficient carrier extraction under zero-bias photovoltaic operation\cite{Sze2006-xc}. Structural and strain characteristics of the double heterostructure are consistent with prior characterization\cite{Atalla2024-yj}, providing a reproducible material platform for the system-level spectroscopic experiments reported above.
 
\FloatBarrier
\subsection{Millimeter-scale Ge$_{0.91}$Sn$_{0.09}$ \texorpdfstring{\pin{}}{p-i-n} photodiode}

The GeSn photodiode investigated in this work was fabricated using a circular mesa architecture with active area diameter up to \qty{1}{\milli\meter}, as shown in Fig.~\ref{fig1}a. This geometry extends previously demonstrated\cite{Atalla2024-yj} vertical GeSn \pin{} devices toward millimeter-scale active areas compatible with broadband spectroscopic measurements. In contrast to high-speed, small-area photodetectors optimized for bandwidth, the present design prioritizes optical collection efficiency and stable operation under low optical power densities.

At millimeter-scale diameters, lateral leakage currents and surface-related generation become increasingly significant due to the enlarged perimeter-to-area ratio and increased junction area \cite{Zhang2022-fz,Ma2016-xh}. To mitigate these effects, a double-mesa configuration was implemented, physically separating the active junction from peripheral conduction paths and reducing lateral current spreading through the buffer layers. This architecture confines the depletion region to the defined active mesa and suppresses edge-related leakage contributions.

Sidewall passivation and dielectric isolation were introduced to minimize surface recombination and reduce Shockley–Read–Hall generation at exposed interfaces. These measures enable stable zero-bias photovoltaic operation across the full \qty{1}{\milli\meter} active area, as reflected by the measured dark current levels reported in Fig.~\ref{fig1}d.

\FloatBarrier
\subsection{Responsivity and I–V measurements}

Current–voltage (I–V) characteristics were measured using a Keithley 4200A semiconductor parameter analyzer connected to a probe station under dark conditions at room temperature.

Spectral responsivity measurements were performed using a Bruker Vertex 80 Fourier-transform infrared (FTIR) spectrometer. The broadband infrared output of the FTIR was directed onto the photodetector under test. The incident beam was mechanically modulated using an optical chopper, and the resulting photocurrent was detected using a Zurich Instruments lock-in amplifier synchronized to the chopper reference frequency to enhance signal-to-noise ratio.

The lock-in amplifier output was recorded as a function of wavelength during the FTIR scan, yielding the photocurrent spectrum. The spectral responsivity $R(\lambda)$ was then calculated by normalizing the measured photocurrent to the calibrated spectral power density of the FTIR source.

\FloatBarrier
\subsection{Detectivity and NEP calculation}

The specific detectivity ($D^*$) and noise-equivalent power (NEP) were calculated from the measured spectral responsivity and dark current under zero-bias photovoltaic operation.

The noise current was estimated by considering both shot noise and thermal (Johnson–Nyquist) noise contributions. The shot noise current was calculated as

\[
i_{\mathrm{shot}} = \sqrt{2q I_{\mathrm{dark}} \Delta f},
\]

where $q$ is the elementary charge, $I_{\mathrm{dark}}$ is the measured zero-bias dark current, and $\Delta f$ is the measurement bandwidth. The thermal noise contribution associated with the shunt resistance $R_{\mathrm{shunt}}$ was estimated as

\[
i_{\mathrm{therm}} = \sqrt{\frac{4 k_B T \Delta f}{R_{\mathrm{shunt}}}},
\]

where $k_B$ is Boltzmann’s constant and $T$ is the absolute temperature. The total noise current was obtained by quadratic summation:

\[
i_n = \sqrt{i_{\mathrm{shot}}^2 + i_{\mathrm{therm}}^2}.
\]

The noise-equivalent power was then calculated as

\[
\mathrm{NEP} = \frac{i_n}{R(\lambda)},
\]

where $R(\lambda)$ is the measured spectral responsivity at the wavelength of interest.

The specific detectivity was computed using

\[
D^* = \frac{\sqrt{A \Delta f}}{\mathrm{NEP}},
\]

where $A$ is the active area of the photodetector. All calculations were performed at 293~K under zero-bias conditions to ensure consistency with the operating mode of the spectrometer platform.

\paragraph{Shunt resistance extraction.}
The shunt resistance $R_{\mathrm{shunt}}$ was extracted from the low-bias slope of the measured dark current–voltage characteristics (Fig.~\ref{fig1}d). Specifically, $R_{\mathrm{shunt}}$ was obtained from the inverse differential conductance around zero bias,
\[
R_{\mathrm{shunt}}=\left(\frac{\mathrm{d}I}{\mathrm{d}V}\right)^{-1}_{V\approx 0}.
\]
To reduce sensitivity to point-to-point noise, the slope $\mathrm{d}I/\mathrm{d}V$ was determined by a linear fit of the dark $I$--$V$ data within a narrow voltage window centered at $V=0$ (typically $|V|\leq V_{\mathrm{fit}}$), and $R_{\mathrm{shunt}}$ was taken as the reciprocal of the fitted slope.

\FloatBarrier
\subsection{NIRscan Nano evaluation module and acquisition parameters}

The spectrometer platform used in this work is the NIRscan Nano Evaluation Module (EVM) developed by Texas Instruments. This compact, single-pixel, microelectromechanical systems (MEMS)-based near-infrared spectrometer uses a digital micromirror device (DMD) for programmable wavelength selection. In its standard configuration, the instrument operates over a spectral range of \SIrange{900}{1700}{\nano\meter}, with a nominal spectral resolution of approximately \SI{10}{\nano\meter}, depending on the micromirror binning configuration.

The system integrates a broadband tungsten--halogen lamp as the illumination source, coupled to the optical path through a sapphire window. Light reflected from the sample is collected and directed through an internal optical assembly comprising a band-pass filter with a cut-on wavelength of \SI{900}{\nano\meter}, a collimating lens, a diffraction grating, a focusing lens assembly, and the DMD. The DMD acts as a programmable spectral filter by selectively directing diffracted wavelength components toward the detector.

The original detector integrated in the module is a standard InGaAs photodiode, which was used to benchmark the system and provide a reference for comparison. In this study, this detector was replaced with a custom GeSn photodiode in order to extend the detectable wavelength range beyond the intrinsic InGaAs cut-off near \SI{1.7}{\micro\meter}. Wavelength scanning and data acquisition were controlled using the Texas Instruments graphical user interface. All measurements were performed under ambient laboratory conditions, with the module powered through USB.

For the InGaAs, Ge, and extended-InGaAs detectors, all scans were performed in the NIR configuration using a spectral pattern width of \SI{2.34}{\nano\meter} and an exposure time of \SI{0.635}{\milli\second}. This acquisition configuration produced 605 spectral sampling points across the selected wavelength range and resulted in a total scan time of \SI{2.1}{\second}. To compensate for the higher noise level of the GeSn detector, measurements in the NIR region were performed using a larger spectral pattern width of \SI{21.08}{\nano\meter} and a longer exposure time of \SI{5.080}{\milli\second}. This configuration increased the collected optical signal per spectral point, produced 492 spectral sampling points, and resulted in a total scan time of \SI{4.6}{\second}.

For the SWIR configuration, acquisitions were performed with both the extended-InGaAs and GeSn detectors using a spectral pattern width of \SI{21.08}{\nano\meter}, corresponding to 492 spectral sampling points. The exposure time was set to \SI{0.635}{\milli\second} for the extended-InGaAs detector and to \SI{5.080}{\milli\second} for the GeSn detector. These acquisition settings resulted in total scan times of \SI{2.1}{\second} and \SI{4.6}{\second}, respectively.

For the GeSn spectra, the multiple-scan averaging option was enabled to improve the signal-to-noise ratio. In this mode, 42 consecutive scans were acquired and averaged by the system, corresponding to an effective acquisition time of approximately \SI{193}{\second} per averaged GeSn spectrum when using the \SI{4.6}{\second} scan configuration.

\FloatBarrier
\subsection{Spectrometer Hardware Development}

To extend the operational wavelength range beyond the conventional InGaAs limit, the internal optical configuration of the spectrometer was modified. The original near-infrared band-pass filter was replaced with a long-pass infrared filter (Edmund Optics) featuring a cut-on wavelength of \SI{1.65}{\micro\meter}. The replacement filter was precision-diced using a micrometer-resolution diamond saw to match the dimensions of the original component, ensuring mechanical compatibility within the existing optical mount and enabling transmission of longer-wavelength components without altering the housing geometry.

The diffraction grating was replaced with a plane ruled grating (model 53-*-770R) with a groove density of \SI{300}{\per\milli\meter} and a blaze wavelength of \SI{2}{\micro\meter}, selected to optimize diffraction efficiency in the short-wave infrared region. Mechanical modifications were required to accommodate the new grating and achieve the diffraction angles necessary for wavelength extension. The original grating mount and adjacent structural elements were carefully removed using precision cutting tools to allow accurate angular positioning of the replacement grating.

To protect the remaining optical components from particulate contamination during modification, a custom protective enclosure was designed and fabricated via 3D printing. This temporary cover isolated the optical cavity during the mechanical adaptation process.

Optical ray-tracing simulations were performed using Zemax to ensure compatibility of the modified components with the fixed internal geometry of the spectrometer. The incident angle on the grating was set to $\alpha_i = -\SI{6.4}{\degree}$, resulting in first-order diffraction at the blaze wavelength with an output angle of $\beta_{m=1} = \SI{45.3}{\degree}$ according to the grating equation. A circular aperture with a diameter of \SI{10}{\milli\meter} defined the collimated beam incident on the grating.

\FloatBarrier
\subsection{Wavelength recalibration and spectral remapping}

The wavelength mapping of the modified spectrometer was recalibrated following detector substitution and optical reconfiguration. An initial reference calibration was performed using a broadband source and a standard InGaAs photodiode with a cut-off wavelength of \SI{1.7}{\micro\meter}. Under these conditions, the detected signal diminished sharply beyond \SIrange{1650}{1700}{\nano\meter}, corresponding to the intrinsic upper limit of the original system response and confirming the baseline spectral mapping.

Following replacement of the InGaAs photodiode with the GeSn detector, the accessible spectral range extended beyond the conventional NIR window. To validate the revised wavelength assignment, a monochromatic laser source at \SI{2330}{\nano\meter} was used as an absolute spectral marker. The laser peak was observed at the expected position within the reconstructed spectrum, confirming accurate remapping of the wavelength axis.

This calibration procedure establishes an effective translation of approximately \SI{740}{\nano\meter} in the operational window, extending the measurable range from \SIrange{900}{1700}{\nano\meter} to \SIrange{1640}{2440}{\nano\meter}. The peak position and full width at half maximum were extracted to quantify wavelength accuracy and spectral resolution in the extended configuration.

\FloatBarrier
\subsection{Spectral resolution calculation of the SWIR spectrometer}

The theoretical resolving power of the diffraction grating was estimated using standard grating theory. For a planar grating, the resolving power is given by $R = mN$, where $m$ is the diffraction order and $N$ is the number of illuminated grooves. Using the grating equation $m\lambda = d(\sin\alpha + \sin\beta)$ and noting that $Nd = W$, where $W$ is the illuminated grating width, the resolving power can be written as

\[
R = \frac{W(\sin\alpha + \sin\beta)}{\lambda}.
\]

Taking into account the grating inclination, the illuminated width was estimated as $W = \Phi \sin\alpha_i$, yielding $W \approx \SI{10073}{\micro\meter}$. At the blaze wavelength of \SI{2}{\micro\meter}, this corresponds to a theoretical resolving power of approximately $R \approx 3.1 \times 10^3$, leading to an intrinsic wavelength resolution of $\Delta\lambda = \lambda/R \approx \SI{0.6}{\nano\meter}$ according to the Rayleigh criterion.

In practice, the effective spectral resolution of the system is limited by the spatial sampling imposed by the digital micromirror device (DMD). The finite number of micromirrors assigned per wavelength channel results in an effective resolution of approximately \SI{10}{\nano\meter} over the investigated spectral range. This value was experimentally confirmed by measuring the full width at half maximum (FWHM) of a \SI{2330}{\nano\meter} laser calibration signal using the GeSn detector at maximum spectral resolution.(Supply material)

The spectrometer operates exclusively in the first diffraction order, and no spectral overlap is expected within the accessible wavelength range. Stray light contributions are minimized by the closed optical architecture and the requirement for collimated illumination on the DMD.

\FloatBarrier
\subsection{Polymer sample preparation}

Representative polymer samples were collected from post-consumer plastic waste and selected to include commonly used commodity plastics: polystyrene (PS), polypropylene (PP), polyethylene terephthalate (PET), and high-density polyethylene (HDPE). Material type was identified based on resin identification codes and manufacturer labeling when available. Polyvinyl (PVC) sheet samples were commercially acquired.  

Samples were manually rinsed with water to remove surface contaminants and dried at room temperature prior to measurement. For spectroscopic characterization, flat regions of each sample were selected to minimize surface curvature effects and ensure consistent optical coupling with the spectrometer illumination.

Dark and black samples were excluded due to their high optical absorption in the investigated spectral range, which significantly reduces transmitted signal intensity and signal-to-noise ratio. All other colors were included without distinction.

All measurements were performed on intact bulk pieces without additional chemical or mechanical treatment, in order to reflect realistic sorting conditions encountered in post-consumer recycling streams. For GeSn measurements, the exposure time and DMD pattern pixel width were increased to compensate for the reduced signal-to-noise ratio associated with extended SWIR operation, resulting in a modest decrease in effective digital spectral resolution while preserving sufficient signal fidelity for polymer discrimination.

\begin{acknowledgement}

The authors acknowledge support from NSERC Canada, Canada Research Chairs, Canada Foundation for Innovation, PRIMA Qu\'ebec, Defence Canada (Innovation for Defence Excellence and Security, IDEaS), the European Union’s Horizon Europe research and innovation program under Grant Agreement No 101070700 (MIRAQLS), and the Air Force Office of Scientific and Research Grant No. FA9550-23-1-0763.

\end{acknowledgement}

% \bibliography{achemso-demo}
\bibliography{bibliography}

\end{document}